\documentclass{emulateapj}

\newcommand {\ergs} {erg s$^{-1}$}
\newcommand {\ergcms} {erg cm$^{-2}$ s$^{-1}$}
\slugcomment{Accepted for publication in ApJ}
\begin{document}

\title{Testing Radiatively-Inefficient Accretion Flow Theory: an
  {\it XMM-Newton} Observation of NGC 3998} 
\author{A. Ptak}
\affil{The Johns Hopkins University, Department of Physics and Astronomy, 
Baltimore, MD 21218}
\author{Y. Terashima}
\affil{Institute of Space and Astronautical Science, 3-1-1 Yoshinodai,
  Sagamihara, Kanagawa 229-8510, Japan}
\author{L. C. Ho}
\affil{The Observatories of the Carnegie Institution of Washington,
813 Santa Barbara S., Pasadena, CA 91101-1292
}
\author{E. Quataert}
\affil{University of California at Berkeley, Astronomy Department,
Berkeley, CA 94720}

\begin{abstract}
We present the results of a 10 ks {\it XMM-Newton} observation of NGC 3998,
a ``type-I'' LINER galaxy (i.e., with significant broad H$\alpha$
emission).  Our goal is to test the extent to which 
radiatively-inefficient accretion flow (RIAF) models and/or 
scaled-down AGN models are consistent with the observed 
properties of NGC 3998.
A power-law fit to the {\it XMM-Newton} spectra results in a power-law
slope of $\Gamma=1.9$ and 2-10 keV flux of $1.1 \times 10^{-11}$
\ergcms, in excellent agreement with previous hard X-ray observations.
% We find that the {\it 
%XMM-Newton} data, when simultaneously fitted with archival 
%{\it ASCA}~ and {\it BeppoSAX} data, show that the 0.1--100 
%keV X-ray spectrum of NGC 3998 is consistent with a simple 
%power-law model with a photon index of $\sim 1.9$. 
The OM UV flux at 2120\AA~ appears to be marginally resolved, with
$\sim 50\%$ of the flux extended beyond 2''.  
%The nuclear component of the flux is $\sim 50\%$ higher
%than the extrapolation of the X-ray power-law.  
The nuclear component of the 2120\AA~ flux is consistent with an
extrapolation of the X-ray power-law, although $\sim 50\%$ of the flux
may be absorbed.
The OM U flux lies
significantly above the X-ray power-law extrapolation and contains
a significant contribution from extragalactic emission.
%This 
%power law also extrapolates to the UV flux determined with 
%the OM. 
The upper-limit for narrow Fe-K emission derived from the {\it XMM-Newton}
spectra is 33 eV (for $\Delta\chi^2=2.7$).  The upper-limit for narrow
Fe-K emission derived 
from a combined fit of the  {\it XMM-Newton} and {\it BeppoSAX}
spectra is 25 eV, which is one of the strictest limits to date for any 
AGN. This significantly rules out Fe-K emission as is 
expected to be observed in typical Seyfert 1 galaxies. The 
X-ray flux of NGC 3998 has not been observed to vary 
significantly (at $>30\%$ level) within the X-ray observations, and only 
between observations at a level of $\sim 50\%$,
which is also in contrast to typical 
Seyfert 1 galaxies.
The lack of {\it any}
reflection features suggests that any optically-thick,
geometrically-thin accretion disk 
must be truncated, probably at a radius of order 100-300 
(in Schwarzschild units).
RIAF models fit the 
UV to X-ray spectral energy distribution of NGC 3998 
reasonably well. In these models the mid-IR flux also constrains the 
emission from any outer thin disk component that might be 
present. The UV to X-ray SED is also consistent with a 
Comptonized thin disk with a very low accretion rate 
($\dot{M} < 10^{-5}\dot{M}_{\rm Edd}$), in which case the lack of Fe-K
emission may be due to an ionized accretion disk.  Accretion 
models in general do not account for the observed radio flux 
of NGC 3998, and the radio flux may be due to a 
jet.  Recent jet models may also be consistent with the nuclear fluxes
of NGC 3998 in general, including the X-ray, optical/UV and mid-IR
bands.  The (ground-based) near-IR
to optical photometric data for the nuclear region of NGC 3998
contains large contributions from extra-nuclear emission.  We also
derive nuclear fluxes using archival HST WFPC2 data, resulting
in meaningful constraints  
to the nuclear SED of NGC 3998 in the optical band. We discuss a possible OM U
band and USNO-B detection of the NGC 3998 ULX.
\end{abstract}

\keywords{X-rays: galaxies --- galaxies: individual (NGC 3998) --- accretion, accretion disks}

\section{Introduction}
Since LINERs (low-ionization nuclear emission-line regions;
Heckman 1980) were first identified as a class the nature of the
source of ionization has been debated, with the two main possibilities
being a starburst or AGN nuclear source.  If an AGN is responsible for
the ionization, the next question is how similar are these AGN to Seyferts
and QSOs.  An interesting development has
been the discovery of broad $H\alpha$ emission in many LINERs \citep{ho97}.
% $\sim XXX\%$ of
%LINERs .  
These LINERs are almost certainly AGN powered,
particularly when they are observed in early-type galaxies where
outflows from massive stars and/or supernovae are not expected to be
relevant.  Early-type galaxies also tend to harbor very massive
central black holes, making broad-line LINERs observed in
early-type galaxies ideal laboratories for studying (potentially) low
accretion rate physics.  Of course, this is assuming that the X-ray emission
is isotropic and scales with accretion rate, however alternatively the
accretion energy may be dissipated anisotropic and/or non-radiatively (i.e.,
in outflows; Di Matteo et al. 2000).

NGC 3998 is an SO galaxy at a distance of 14.1 Mpc \citep{to01}. Its
H$\alpha$ line 
contains a broad  component with 37\% of the H$\alpha$ + [N II]
flux and a FWHM of $\sim 2000 \rm \ km \ s^{-1}$ \citep{ho97}.  
{\it ASCA} and {\it BeppoSAX} observations find that the 2-10 keV luminosity of
% Adjusted 7/23/03 based on new distance of 14.1 Mpc
NGC 3998 is $\sim 3 \times 10^{41}$~\ergs \citep{pe00, te02},
and the 0.1-100 keV luminosity is $\sim 1 \times 10^{42}$~\ergs.  
The mass of a putative black hole in the nucleus of NGC 3998
is estimated to be $\sim 10^{9} M_{\odot}$, giving $L_X/L_{Edd}
\sim 1 \times 10^{-5}$, or well into the radiatively-inefficient 
accretion flow (RIAF) regime.  The main goal of this paper is to contrast
NGC 3998 with ``normal'' Seyfert 1 galaxies and to test
the applicability of current RIAF models to the observed properties of
NGC 3998.
%A general property of these types of
%accretion flows is a low particle density which in turn results in a
%low radiative time scale compared with dynamical time scales.  
%This
%results in accretion flows that are optically thin
%and accordingly radiate via bremsstrahlung rather than blackbody
%processes.  
%In contrast the optically-thick, geometrically-thick
%``$\alpha$'' accretion  disks thought to be present in typical AGN
%produce a blackbody spectrum that is thought to peak in the blue-UV
%bandpass and is responsible for the ``blue bump'' observed in AGN.

%The primary emphasis of this paper is on 
We present the results of a recent
{\it XMM-Newton} observation of NGC 3998, analyzed in conjunction with
archival X-ray data.  Previous X-ray observations showed
that no Fe-K line is present in the X-ray spectrum from NGC 3998, and
here we tighten this constraint, which places limits on the accretion
geometry in NGC 3998.  Additionally we will use {\it XMM-Newton} Optical
Monitor data along with other multi-wavelength data to compare the
spectral energy distribution (SED) of NGC 3998 with those predicted by
various accretion models.

\section{Data Reduction}
The parameters of X-ray observations are listed in Table~\ref{tab1}.
The {\it ASCA} data was reduced as described in \citet{pt99} and \citet{te02}.  
The {\it BeppoSAX} data were reduced by the XSELECT and FTOOLs in HEASOFT
5.2 from the cleaned event files obtained from the {\it BeppoSAX} Science
Data Center. LECS and MECS spectra of NGC 3998 were extracted from a
circular region with a 4 arcmin radius. Background spectra were taken
from an annular region around NGC 3998.
Here we present CCD spectra from
the three EPIC detectors (two MOS and one PN) and the Optical Monitor
U-band and UV 
(UVW2 filter) images.  
%NGC 3998 is unresolved in the X-ray and UV
%images.  
The {\it XMM-Newton} data was reduced using XMMSAS version
5.4.1 with the use of XAssist, a software package that performs the
initial steps of X-ray data analysis \citep{pt02}.  The XMM PN and MOS spectra
were extracted from 37'' circular regions (derived from fitting a Gaussian
surface brightness model to the X-ray images).  
We compared the radial profile of the mos1 image (chosen since the
target was not close to a ccd boundary in that case) to a SciSim
simulation using the spectral model (see \S \ref{specsec}) and found
that the X-ray emission from NGC 3998 is dominated by an unresolved 
source.  The only other significant source in the vicinity of NGC
3998 is NGC 3998 X-1 \citep{ro00} which is
discussed in \S\ref{ulxsec}. 
%The nearest source beyond the RC3 $d_25$ extent of NGC 3998
%is 
The OM data was reprocessed with XMMSAS and the images and source lists
produced by the tool ``omichain'' were used in this paper.  The OM spatial
resolution depends on photon energy and the FWHM of the point-spread function
(PSF) is $1.6''$ and  $1.9''$ in the U and UVW2 bands 
({\it XMM-Newton} User's Guide).  
%The spectra from the individual ASCA SIS and GIS detectors were
%combined.
All spectra were binned to 20 counts/bin to allow the use of the
$\chi^2$ statistic. 

%\clearpage

\begin{deluxetable*}{llll}
%\tablestypesize{\scriptsize}
\tablecaption{Observation Parameters \label{tab1}}
\tablehead{
\colhead{Satellite} & \colhead{Bandpass (keV)} & \colhead{Date} &%
\colhead{Exposure Time (ks)} 
}
\startdata
{\it ASCA} & 0.4-10.0 & 05/10/94 & 40 \\
{\it BeppoSAX} & 0.1-100.0 & 06/29/99 & 24 (LECS), 77 (MECS), 38 (PDS) \\
{\it XMM-Newton} & 0.2-10.0 & 05/09/01 & 12 (MOS), 9 (PN)
\enddata
\end{deluxetable*}

%\clearpage

\section{Results}
\subsection{Spectral Analysis \label{specsec}}
\begin{figure}
%\plotone{n3998_xmmhighbgdpn0-4_asca_sax_pl_ascanh_06nov02_marked.ps}
%\plotone{n3998_xmm_pl_22jul03.ps}
\plotone{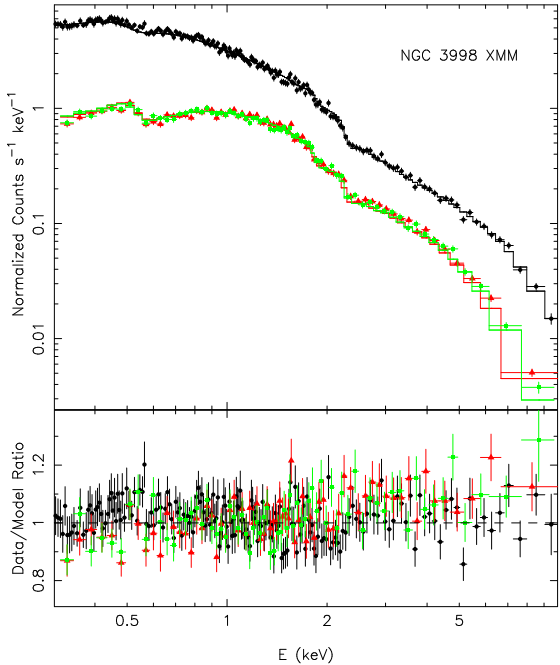}
%\plotone{f1_bw.eps}
\caption{Power-law fit to the {\it XMM-Newton} spectra.  The data and best-fitting
model are shown in the top panel and the ratio of data to model is shown in 
the bottom panel.  The PN, MOS1 and MOS2 data and residuals were
marked with filled circles (black), filled triangles (red) and filled squares
(green), respectively.
\label{xmmplfig}}
\end{figure}

We fit the {\it XMM-Newton} (CCD) spectra of NGC 3998 with a simple absorbed
power-law model, allowing the overall normalization of the model fit
to each detector to be independent.   This resulted in a good fit with
$\Gamma \sim 1.9$, consistent with the previous {\it ASCA} and
{\it BeppoSAX} observations as shown in Table \ref{specfittab}.  
Note that while the three observations resulted in very close
agreement for the power-law slope, the column density measurements varied from
$3.3 \times 10^{20} \rm \ cm^{-2}$ (or a factor of $\sim 3$ greater than the
Galactic value; Murphy et al. 1996) to $8.0 \times 10^{20} \rm \
cm^{-2}$.  Given the calibration uncertainties below 0.5-1.0 keV for
these detectors it is not clear to what extent these variations are
significant.  We show the power-law fit to the {\it XMM-Newton} data in
Figure~\ref{xmmplfig} (the {\it BeppoSAX} and {\it ASCA} data are plotted in
Pellegrini et al. 2000 and Terashima et al. 2002).  The {\it ASCA} data also
resulted in a flux $\sim 30\%$ lower than the {\it BeppoSAX} and {\it XMM-Newton}
measurements (see \S \ref{varsec}). Note that the PDS flux was
$\sim 45\%$ higher the mean of the other flux measurements, however
the bandpass of the PDS 
data was $\sim 10-100$ keV so the 2-10 keV flux is based on an
extrapolation.  We also tried restricting the {\it BeppoSAX} data to only
include data 
below 10 keV (i.e., excluding the PDS data) and found no significant
impact on the fit parameters.  Power-law fit parameters from a fit to
the {\it XMM-Newton} and {\it BeppoSAX} data simultaneously, with the normalization of each data set allowed to
vary, is shown in Table \ref{specfittab}, and $\chi^2$ only increased
by 3 when the normalization of the 
PDS component was fixed result in the same 2-10 keV flux as the MECS
flux (i.e., the PDS data systematically deviate from the power-law fit
at a confidence of only $\sim 90\%$).  

%We also allow the column density
%to vary separately for the ASCA detectors since there are 
%known low-energy calibration uncertainties
%(\url{http://asca.gsfc.nasa.gov}).  
%%This resulted
%%in a good fit, shown Figure~\ref{xmmplfig}, with best-fitting parameters
%%and errors 
%%(based on $\Delta \chi^2 = 2.7$) listed in 
%%Table~\ref{specfittab}.  The photon index of the power-law 
%%was found to be 1.87, typical of Seyfert 1 galaxy spectra \citep{mu93}.  
%%The 2-10 fluxes determined using the individual
%%detectors were typically within 20\% of the mean, except in the cases
%%of the ASCA and {\it BeppoSAX} PDS detectors.  The ASCA fluxes were
%%$20-25\%$ lower than and  Since the ASCA detectors agree with each other to within
%%5\%, it is likely that nucleus of NGC 3998 varied at the $\sim 25\%$
%%level between the the ASCA observation in 1994 and the {\it BeppoSAX} and
%%{\it XMM-Newton} observations in 1999 and 2001, respectively.  
%%

There are no systematic residuals suggesting additional spectral
features might be present.  We do not find evidence for any edges
suggesting that an ionized absorber is not present (note that the edges
discussed in Pelligrini et al. were detected at less than the
3$\sigma$ level).
The RGS data do not
have high enough signal-to-noise to constrain spectral features.
There is a slight excess of counts in the 0.65-0.7 keV region but this
feature is not statistically significant.  We also tried a plasma plus
power-law fit \citep{pt99} to the {\it XMM-Newton} spectra and found a
significant improvement in $\chi^2$ ($\chi^2/dof$ = 1406/1334 compared
with 1423/1338 for the power-law fit).  An absorption component
applied to the whole spectrum resulted in $N_H = 4.6 (1.0-5.8) \times
10^{20} \rm \ cm^{-2}$ and a separate absorbed applied to just the
power-law component resulted in an upper limit of $4.9 \times 10^{20}\
\rm cm^{-2}$. 
The temperature of the plasma component was 0.24 (0.18-0.43) keV, and
the abundance was 0.016 ($<0.024$).  Note that the bulge velocity
dispersion in NGC 3998 is 304 km s$^{-1}$ \citep{mc95}, implying
a virial temperature of $\sim 0.6$ keV.  Since NGC 3998 is an
early-type galaxy and the inferred abundance from this fit is low, it
is likely that this component is due to virial heating of the ISM
(rather than starburst emission as is often observed in LLAGN).
The power-law slope in this  
case was 1.84 (1.84-1.89), unchanged within the errors from the simple
power-law fit. 
The thermal component obviously
contributes to the spectrum most significantly at the lowest energies,
with a 0.3-0.7 keV flux of $4 \times 10^{-13}$ \ergcms, or $\sim
13\%$ of the total 0.3-0.7 keV flux.  However note that in this band
the flux estimates of the 3 EPIC detectors differ by $\sim 8\%$ and
therefore this component should be treated with caution.

In the {\it ASCA} GIS spectrum
there is a small excess of counts near 6.4 keV which results in a
``detection'' of an Fe-K line.  However the addition of a narrow
Gaussian line ($\sigma = 0.01$ keV%, less than the instrumental
				%resolution of CCD detectors
) to the {\it ASCA} power-law fit only reduces
$\chi^2$ by 3.7 
\citep{te02}, which is only significant at a confidence of $\sim
96\%$ based on the f-test.  
In Table~\ref{ewtab} we
list the upper-limits obtained for narrow  line emission
with the energy fixed at 6.4 keV.  
Not surprisingly, the limit obtained when {\it {\it ASCA}} data is included
in the fit is {\it higher} than when the {\it XMM-Newton} and 
{\it BeppoSAX} data are fit alone.  In this table we list 90\% errors
for one interesting parameter ($\Delta \chi^2=2.7$), which are
appropriate since here there is only one parameter of interest (the
normalization of the line), however we also give $\Delta  \chi^2=4.6$
errors for comparison with 2-parameter errors in the literature (or
alternatively 97\% 1-parameter errors).  
%This is due to the presence of a
%faint Fe-K line detected in the {\it ASCA} data alone \citep{te02},
%although the addition of a narrow Gaussian line to the power-law fit
%of the {\it ASCA} data only reduced $\chi^2$ by 3.7 (with the f-test giving
%a significance of 96\%).  
We also attempted to fit a disk line model
(with the energy fixed at 6.4 keV, inclination fixed at 0$\degr$,
power-law emissivity slope fixed at 
-2 and the inner and outer radii fixed at 5 and 500 in Schwarzschild units), 
and obtained upper-limits
%(for $\Delta \chi^2 = 4.6$) of 34 eV and 43 eV respectively with and
%without
% from n3998_xmm+sax+asca_pldisk_asnhfree_22sep03, 
% n3998_xmm+sax_pldisk_snhfree_22sep03.xcm
(for  $\Delta \chi^2$ = 2.7) of 33 eV and 42 eV, without
and with the {\it ASCA} data included, respectively.
%although there is an excess in the 0.6-0.7 keV region.  This feature
%is fit will be a Gaussian line with centroid energy = 0.67 keV,
%physical width = 30 eV, and EW =
%18 eV.  However, this feature is only significant at the XXX level.

%\clearpage

\begin{deluxetable*}{llllll}
\tabletypesize{\scriptsize}
\tablecaption{Power-Law Fits to X-ray Spectra \label{specfittab}}
\tablehead{
\colhead{Detector} & 
\colhead{$N_H$} &
\colhead{$\Gamma$} & 
\colhead{$F_{\rm 2-10\ keV}$} & 
\colhead{$\chi^2/dof$} & 
\colhead{Reference} \\
& \colhead{($10^{20} \rm \ cm^{-2}$)} &
& \colhead{($10^{-11}$ \ergcms)}
}
\startdata
%ASCA & 8.8 (7.8-9.8) & 1.89 (1.87-1.92) & 0.80 & 819.4/836 & This work
%\tablenotemark{1}\\ % /bit/data1/llagn/manual/n3998/spectra_22jul03/n3998_asca_pl_wabs.log
{\it ASCA} & 8.0 (6.9-9.2) & 1.89 (1.86-1.92) & 0.80 & 310.9/281 & This work\tablenotemark{1}\\  %/bit/data1/llagn/manual/n3998/spectra_22jul03/asca/n3998_asca_pl.log
{\it BeppoSAX} &  5.4 (3.5-7.8) & 1.87 (1.83-1.90) & 1.2 & 222.8/227 & This
work\tablenotemark{2} \\ %/bit/data1/llagn/manual/n3998/spectra_22jul03/n3998_sax_pl_new.log
{\it BeppoSAX}, $<$10 keV & 5.1 (3.4-7.5) & 1.86 (1.82-1.90) & 1.1 & 207.9/210 & This
work \\  % /bit/data1/llagn/manual/n3998/spectra_22jul03/n3998_sax_pl_lt10kev_new.log
{\it XMM-Newton} & 3.3 (3.0-3.6) & 1.88 (1.87-1.90) & 1.1 & 1422.6/1338 & This work\\
{\it XMM-Newton} + {\it BeppoSAX} & 3.3 (3.0-3.6) & 1.88 (1.87-1.89) & 1.1 & 1652.5/1567 & This work\\ % n3998_xmm+sax_pl_22jul03.log

\enddata
\tablenotetext{1}{See also \citet{te02}}
\tablenotetext{2}{See also \citet{pe00}}
\tablecomments{Errors are given in parenthesis for the 90\% confidence
  interval assuming one interesting parameter (i.e., $\Delta\chi^2 = 2.7$).}
\end{deluxetable*}

\begin{deluxetable*}{lll}
\tabletypesize{\scriptsize}
\tablecaption{Fe-K Equivalent Width Limits \label{ewtab}}
\tablehead{
\colhead{Detector(s)} & 
\colhead{EW, $\Delta \chi^2$=2.7} &
\colhead{EW, $\Delta \chi^2$=4.6}
}
\startdata
{\it XMM-Newton} &  $<$ 32.7 & $<$41.4 \\
{\it BeppoSAX} & $<$38.2 & $<$56.0\\
{\it ASCA} & 85.5 (11.7-159) & $<$186\\ % asca/n3998_asca_plga_esfrozen_new.xcm
{\it XMM-Newton} + {\it BeppoSAX} & $<$25 eV & $<$ 33 eV\\
{\it XMM-Newton} + {\it BeppoSAX} + {\it ASCA} & $<$33 & $<$40\\
\enddata
\tablecomments{In all fits the line energy and physical width were fixed at
6.4 keV and 0.01 keV (i.e., a narrow line), respectively.}
\end{deluxetable*}

%%%\begin{deluxetable}{llll}
%%%\tabletypesize{\scriptsize}
%%%\tablecaption{Spectral Fits to X-ray Spectra \label{specfittab}}
%%%\tablehead{
%%%%& \multicolumn{2}{c}{{\it XMM-Newton} + {\it BeppoSAX}} & 
%%%%\multicolumn{2}{c}{{\it XMM-Newton} + {\it BeppoSAX} + {\it ASCA}} \\ 
%%%& & \colhead{{\it XMM-Newton} + {\it BeppoSAX}} & \colhead{{\it XMM-Newton} + {\it BeppoSAX} + {\it ASCA}}\\
%%%\colhead{Model} & \colhead{Parameter} & \colhead{Value} &
%%%\colhead{Value}
%%%}
%%%\startdata
%%%% n3998_xmmhighbgdpn0-4_sax_pl_02oct02.log|
%%%Power-law & $N_H (10^{20} \rm \ cm^{-2})$ & 
%%%3.4 (3.1-3.7) & 3.5 (3.2-3.8), {\it ASCA} 8.2 (7.3-9.1)\\
%%%& $\Gamma$ & 1.87 (1.85-1.89)  & 1.88 (1.86-1.89) \\
%%%& $\chi^2/dof$ & 1617/1504 & 2437/2341 \\
%%%& 2-10 keV Flux\tablenotemark{*} & 1.1 & 1.0 \\
%%%Power-law + Gaussian & 
%%%%$N_H (10^{20} \rm \ cm^{-2})$ & 
%%%%3.4 (3.1-3.7) & 3.5 (3.2-3.8), {\it ASCA} 8.2 (7.3-9.1)\\
%%%%& $\Gamma$ & 1.87 (1.85-1.89) & 1.88 (1.86-1.89)\\
%%%%& 
%%%Fe-K Flux\tablenotemark{**}  & 0. ($<2.7$) & 0.5 ($<4.0$)\\
%%%& Fe-K EW (eV) & 0 ($<24$) & 2 ($<33$)\\
%%%%& $\chi^2/dof$ & 1617(1503) & 2437/2340 \\
%%%\enddata
%%%\tablenotetext{*}{in units of $10^{-11}$~\ergcms}
%%%\tablenotetext{**}{in units of $10^{-14}$~\ergcms}
%%%\tablecomments{The power-law model photon index, column density and
%%%  flux fit parameters were not affected by the addition of the
%%%  Gaussian line component to the fit.}
%%%\end{deluxetable}
%%%

%\clearpage

\begin{figure*}
%\plottwo{n3998_xmm+sax+asca_plga_esfrozen_asnhfree.ps}{n3998_xmm+sax+asca_plga_esfrozen_asnhfree_5.0-7.5kev.ps}
\plottwo{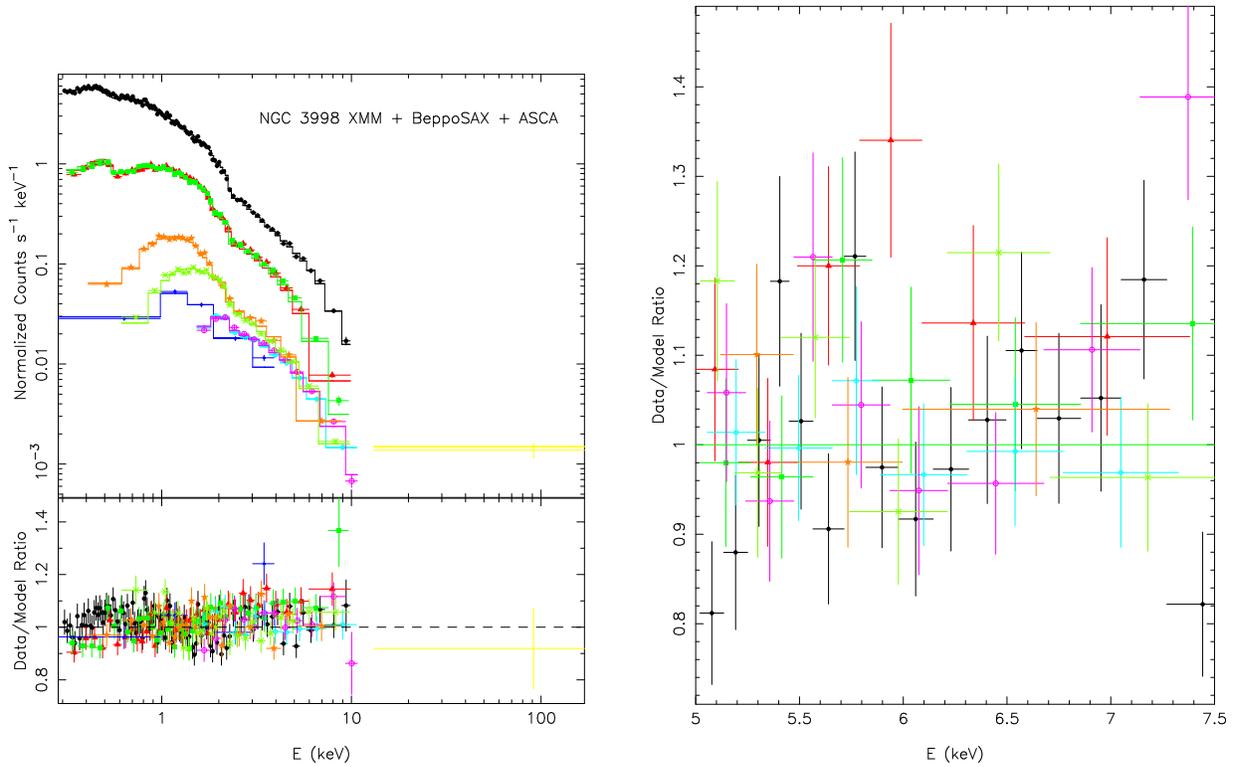}{f2b.eps}
%\plottwo{f2a_bw.eps}{f2b_bw.eps}
\caption{Power-law fit to the {\it XMM-Newton}, {\it BeppoSAX} and {\it ASCA} spectra.
  The normalization of each data set was allowed to vary
  independently.  The column density of a cold absorber was allowed to
  vary independently for each observation (but was fixed to be the
  same for each {\it XMM-Newton} detector, for example).  
  The data are marked as follows: XMM PN, MOS1, MOS2 = filled circles,
  filled triangles, and filled squares; {\it ASCA} SIS = filled star,
  GIS = crosses; {\it BeppoSAX} LECS, MECS2-3 = circles of increasing size,
  PDS = unmarked (the only data point beyond 10 keV).  The data and
  best-fitting model are shown 
  in the top panel and the ratio of data to model is shown in  
  the bottom panel of the left plot, and the right plot shows only the
  data/model ratio in the 5.0-7.5 keV region.
  \label{allplfig}}
\end{figure*}

%\clearpage

%Power-law fit, upper-limit to Fe-K, upper-limit to warm absorber
%edges, check for ionized Fe edge possibly detected in SAX data.

\subsection{Optical Monitor Data} 
\begin{figure*}
%\plottwo{n3998_u_nucl.ps}{n3998_uvw2.ps}
\plottwo{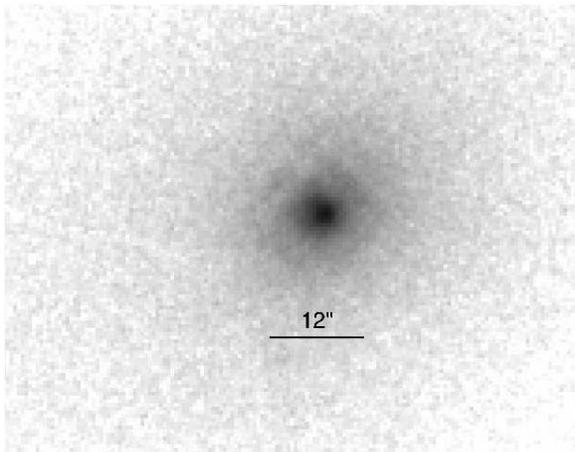}{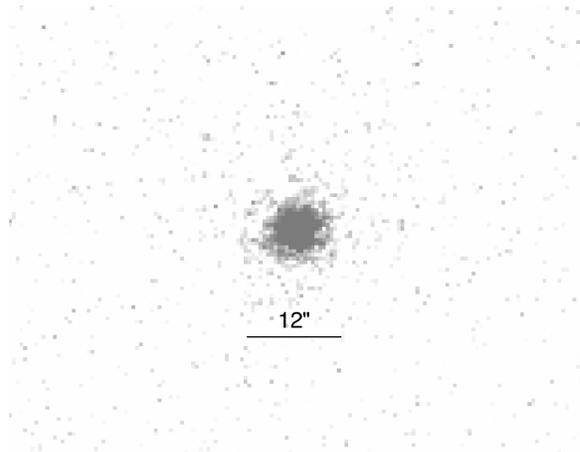}
\caption{OM images of the central region of NGC 3998 in the U (left)
  and UVW2 (right) bands \label{omfig}}
\end{figure*}

The OM detector exposed five 1000 s images in the U and UVW2
filters, for a net exposure time of 5000 s in each band (additional
exposures were taken of flanking fields which will not be discussed
here).   The central regions of the U and UVW2 images are shown in
Figure \ref{omfig}.  In order to properly model the surface brightness
distributions for the two filters we also required model PSFs (there
were no stars sufficient for this purpose in our observations).
We initially produced model PSFs using SciSim.  However, the current
version of SciSim 
(3.0.0) does not specifically model the PSF-dependence of the
individual filters, and also only models the PSF as a single Gaussian.  
Nevertheless, the resultant PSF FWHM was $\sim
2.1$'' which is similar to that expected for the UVW2 filter and
therefore we proceeded with this PSF model for that filter.  In the
case of the U band filter we instead produced a model PSF by
generating double Gaussian images and adjusting standard deviations of
the two Gaussians and their relative normalization until the encircled energy
function (i.e., the integrated
radial profile) matched the values given in the OM calibration file
OM\_PSF1DRB\_0006.CCF.  In the surface brightness fitting described
below we model the nuclear point-source component with a Gaussian
whose standard deviation ($\sigma$) is permitted to be a free
parameter.  The fitted value of $\sigma$ would then
represent some combination of telescope jitter during the observation
and also any adjustment to the PSF due to the inadequacies of our PSF model.

For UVW2 data, the
XMMSAS tool ``omichain'' 
produces OM point source lists based on a 6'' aperture and background
taken from an aperture with radii 7-12''.  The nuclear 2120\AA~
magnitude was 15.1 which corresponds to a flux density of $4.9 \times
10^{-15}$~\ergcms$\AA^{-1}$ (based on a Vega zero point and flux of 
0.025 and $5.39 \times 10^{-9}$~\ergcms$\AA^{-1}$\footnote{all OM flux
conversions discussed here are based on the calibration given at\\
http://xmm.vilspa.esa.es/sas/documentation/watchout/uvflux.shtml}). 
The UVW2 images were fit with a double Gaussian model (with one
Gaussian intended to represent any diffuse flux) from
which it was determined that $\sim 50\%$ of the UV flux from NGC 3998
is extended over a $\sim 6''$ (FWHM) scale (the individual fits
typically resulted in Gaussian $\sigma$ values of $\sim 1$'' for the
nuclear component and $\sim 3-4$'' for the diffuse component).  
Our spatial analysis implies that the flux density from within a 2''
aperture was    
$\sim 2.5\times 10^{-15}$~\ergcms$\AA^{-1}$. Similarly, 
integrating the Gaussian fit for the nuclear component results in a
count rate of $\sim 0.5$ counts s$^{-1}$, which implies a flux
density of $\sim 2.4 \times 10^{-15}$~\ergcms$\AA^{-1}$ (using a count
rate to flux conversion factor of $4.871 \times
10^{-15}$~\ergcms$\AA^{-1}$ counts$^{-1}$ s).  
This result is in
fairly good agreement with the values determined spectrally from IUE data
in \citet{re92}, $1.5 \times 10^{-15}$~\ergcms$\AA^{-1}$ and $1.2
\times 10^{-15}$~\ergcms$\AA^{-1}$ for the unresolved and extended
(FWHM $\sim 10''$) continuum components in the 1950-2155\AA~ region. 
Given the uncertainties (particularly in the PSF model), we note that
the actual nuclear 
flux density most likely lies in the range of $2.5-5.0  \times
10^{-15}$~\ergcms$\AA^{-1}$.  We therefore assume a flux of $3.7 \times
10^{-15}$~\ergcms$\AA^{-1}$ with a 50\% uncertainty.
However, this value is a factor of $\sim 2-4$ lower
than the FOC lower limit ($1.0 \times 10^{-14}$~\ergcms$\AA^{-1}$)
determined by \citet{fa94}.  Since the FOC UV 
data was overexposed in that observation our analysis will
be based on the OM flux, and we note that it is possible that HST
observed a flare in the UV bandpass.  

The U band magnitude was
determined from a 3'' aperture (using the XMMSAS tool ``omsource''
which allowed for smaller source radii in the case of U band images).
%, with the background being derived
%locally, which we then rescaled to 2'' using the ratio of counts within radii
%of  2'' and 6''.
The U band magnitude was 14.3, which corresponds to a flux
of $6.1 \times 10^{-15}$ \ergcms$\AA^{-1}$ 
(with the Vega zero-point and flux being
0.025 and $3.16 \times 10^{-9}$~\ergcms$\AA^{-1}$ flux).  
%The U band morphology is
%mostly symmetric.% and accordingly
%the galactic flux is not perfectly subtracted.  
The U band flux contains a contribution from the extra-nuclear emission but
can of course nevertheless be considered to be an upper-limit
to the nuclear U-band flux of NGC 3998. We fit the central images with
a model that consisted of a Gaussian (again to represent a point source
including jitter) and a de Vaucouleurs profile ($\propto \exp
-7.669[(\frac{r}{r_e})^{1/4} - 1.)]$, where $r_e$ is the scale length
of the profile.  These fits resulted in $r_e$ values of $\sim 30-40$'',
although the precise values are not particularly meaningful given the
uncertainty in the PSF and the fact that we only fit the central 30'' of
the images since we are mainly interested in the determining the
nuclear flux.   The Gaussian $\sigma$ values were $\sim
1.5''$, with the Gaussian component contributing 40-60\% of the flux
within a radius of 3''.  Here the count rate integrated from the
nuclear Gaussian component is $\sim 11$ counts s$^{-1}$, which results in
a flux density of $2.1 \times 10^{-15}$~\ergcms$\AA^{-1}$.  
The U band HST FOC image was also overexposed, and \citet{fa94} report
a lower-limit of $4.2 \times 10^{-16}$\ergcms$\AA^{-1}$ which is consistent
with our flux.

The current flux calibration uncertainty for OM U and UV magnitudes is
$\sim 10\%$ (U.S. {\it XMM-Newton} Guest Observer Facility, private communication).
To check the 
calibration of the U band magnitudes, we computed the 
the (total) count rate of NGC 3998 and another galaxy in
the U band 
FOV, NGC 3990, and derived U band magnitudes of 11.92 and 13.94. These
values can be compared to the RC3 U band magnitudes of 12.13 and 13.81,
respectively, or differences of 0.21 and 0.13. This implies that the
flux calibration is indeed accurate to at least $\sim 10\%$.  As with
the UV flux, the nuclear U band flux the uncertainty is dominated
by the contribution of the extra-nuclear component, and
therefore we assume a value of $3.0 \times 10^{-15}$~\ergcms$\AA^{-1}$
(50\% of the aperture photometry value) with a $\sim 30\%$ uncertainty
(i.e., with 1$\sigma$ being the offset to the model nuclear flux value
discussed above).   

%We estimate the uncertainty in U and UVW2 flux conversion factors to be on the
%order of $\sim 20\%$.
%The uncertainty on the OM zero-points
%is a few percent ({\it XMM-Newton} calibration document CAL-TN-0019-1-0).
%http://xmm.vilspa.esa.es/docs/documents/CAL-TN-0019-1-0.ps.gz

%\clearpage

%\clearpage

In order to supplement the OM data with additional optical photometric
points, we derived nuclear fluxes from the HST WFPC2 F547M
(u2uh0505t, 200 second
exposure) and F791W
(u2uh0507t, 80 second exposure) observations of NGC 3998.  For both images
the net count rate was computed  within a 0.1'' circular aperture (with the
background estimated from an annulus at 2.5-2.6''). The count rates
were converted to fluxes using the PHOTFLAM FITS keywords ($2.96
\times 10^{-18}$~\ergcms$\AA^{-1}$ and $7.69 \times
10^{-18}$~\ergcms$\AA^{-1}$
for F791W and F547M, respectively). 
This resulted in flux densities of $7.4 \times 10^{-16}$~\ergcms $\AA^{-1}$ and
$7.9 \times 10^{-16}$~\ergcms $\AA^{-1}$ at 7872\AA~ and 5484\AA,
respectively. We produced model PSFs for these 
images using TinyTim and fitted the central 3'' (i.e., a $64 \times
64$ pixel image)  with a Gaussian plus de Vaucouleurs model.  This
resulted in nuclear  flux density estimates of $6.6 \times
10^{-16}$~\ergcms $\AA^{-1}$ and $7.6 \times 10^{-16}$~\ergcms
$\AA^{-1}$, which do not differ by more than $\sim 10\%$ from values
obtained above from aperture photometry, and we assume an error of
$10\%$ on these fluxes.  
 
%This
%resulted in flux densities of $1.5 \times 10^{-26}$\ergcms Hz${-1}$ at 
%7872\AA and $7.9\times 10^{-27}$\ergcms Hz${-1}$ at 5484\AA.  

\subsection{Variability \label{varsec}}
\begin{figure}
%\plotone{n3998_lt_lc_24sep03.eps}
\plotone{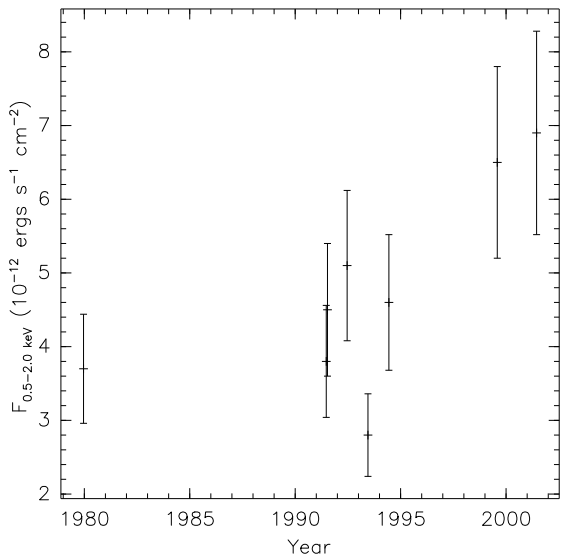}
\caption{Long-term 0.5-2.0 keV light curve for NGC 3998 \label{lcfig}.
  All points are plotted with errors at the level of 20\% ($\sim
  2\times$ the variation in flux calibration typically observed
  between X-ray detectors. See Snowden (2002).}
\end{figure}

No statistically-significant short-term variability has been observed
in the X-ray data, except for a marginally-significant, $\sim 12\%$ scale
variation observed in the MECS data \citep{pe00}.  In
Figure~\ref{lcfig} we show
the long-term 0.5-2.0 keV light curve of NGC 3998.  These fluxes were
derived from the {\it ASCA}, {\it XMM-Newton}, and {\it BeppoSAX} data discussed in this
paper as well as archival ROSAT PSPC and HRI data (in which we
converted the count rate to the 0.5-2.0 keV flux using a power-law
with a slope of 1.88).  The Einstein data point from 1979 is from \citet{bu97}
converted to the 0.5-2.0 keV bandpass.  We chose 0.5-2.0 keV bandpass
since it is within the energy ranges of these detectors, and, as shown
in \S\ref{specsec}, there is a negligible contribution from hot gas to
the spectrum (as would be expected since NGC 3998 is not a
star-forming galaxy).
From this plot, it appears that
the flux of NGC 3998 was fairly constant during the measurements but
may have been a factor of $\sim 50\%$ higher during the {\it BeppoSAX} and
{\it XMM-Newton} observations.  The Gina 2-10 keV flux for NGC 3998 was
also high, by a factor of $\sim 1.9$ ($\sim 1.3$) relative to the {\it ASCA}
({\it BeppoSAX}) fluxes, although it should be kept in mind that Ginga was not an 
imaging telescope and had a FOV of $\sim 1\degr$ \citep{aw91}.  Note
that X-ray binaries are not expected to be a significant contributor to
the X-ray luminosity of NGC 3998 since typical normal galaxy
luminosities tend to be $\lesssim 10^{40}$ \ergs \citep{da92}, in
contrast to $\sim 3 \times 10^{41}$ \ergs for NGC 3998.  More
specifically, the B-band luminosity of NGC 3998 is $\sim 1.6 \times
10^{43}$\ergs (based on $B_T = 11.6$ from RC3), which implies a
0.5-4.5 keV luminosity of $\sim 1.6 \times 10^{39}$ \ergs using the
correlation in \citet{da92}. 

%\clearpage

%\clearpage

\subsection{NGC 3998 X-1 \label{ulxsec}}
\begin{figure*}
%\plottwo{n3998_pn_x-1_nobgd.ps}{n3998_u_marked_150dpi.ps}
\plottwo{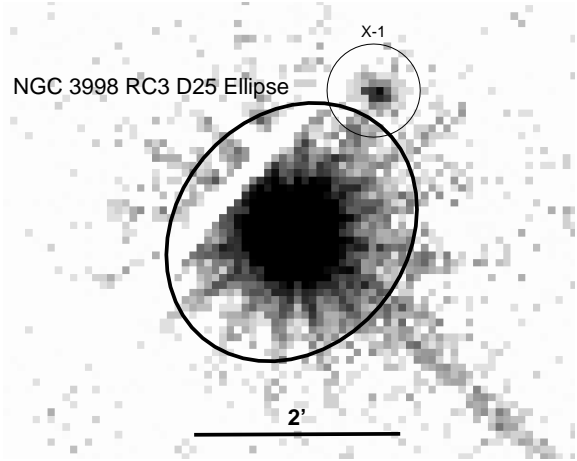}{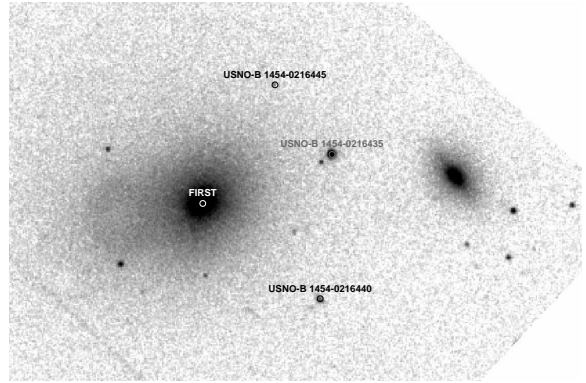}
\caption{NGC 3998 PN (left) and U (right) images of NGC 3998.  The PN
  image is marked with the RC3 $d_{25}$ ellipse for NGC 3998 as well
  as the position of NGC 3998 X-1.
  The U band image shows the
  FIRST position for NGC 3998 along with USNO-B 1454-0216435 and
  USNO-B 1454-0216440 after shifting by $\sim 1''$ in order to align
  with the nucleus of NGC 3998 and two stars in the field.  Also shown
  is the position of USNO-B 1454-0216445 which aligns with a U band
  source that may be associated with NGC 3998 X-1.  The {\it XMM-Newton}
  X-ray position for X-1 is $11\degr 57^m 50^s, 55^h 28^m 34^s$
  J2000.  The position of USNO-B 1454-0216445 is $11\degr 57^m
  50.27^s, +55^h 28^m 34.9^s$ (offset 2.1'' from the X-ray position).
 \label{pn_x-1_fig}}
\end{figure*}

\begin{figure}
%\plotone{ulx_xmm_pl_30sep03.ps}
\plotone{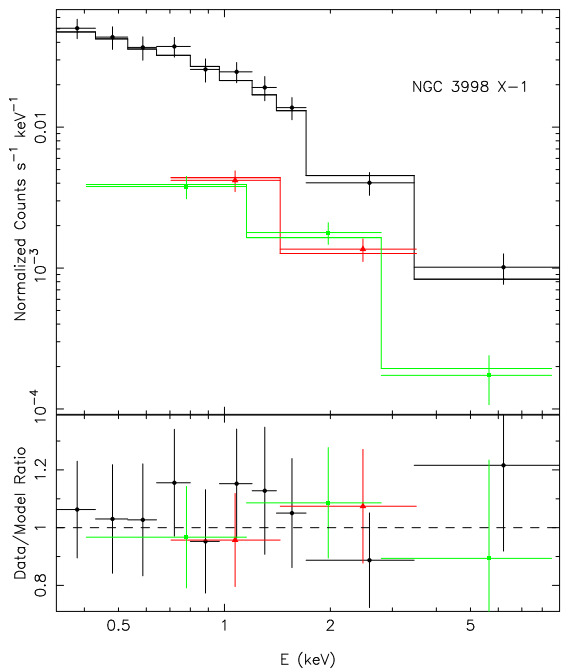}
%\plotone{f6_bw.eps}
\caption{Power-law fit to the PN, MOS1 and MOS2 data (marked with
  filled circles, triangles, and squares, respectively) for NGC 3998
  X-1.\label{ulx_pl}}
\end{figure}

NGC 3998 X-1 \citep{ro00} is an extra-nuclear X-ray source just
outside of the RC3 $d_{25}$ ellipse of NGC 3998.  In Figure \ref{pn_x-1_fig} we
show the PN image of the central $\sim 4'$ of NGC 3998 with the position
of X-1 marked.  X-1 was detected with 441, 73, and 84 net counts by the
PN, MOS1 and MOS2 detectors, respectively, in 27.5'' regions.
Background spectra were extracted from circular regions of the same
size in a source-free region away from the ULX and NGC 3998.
A power-law fit resulted 
in $N_H = 0.5 (<5.7) \times 10^{20} \rm \ cm^{-2}$, 
$\Gamma = 1.8 (1.6-2.1)$ and
$\chi^2$/dof = 25.9/32 (see Figure
\ref{ulx_pl}).  The 0.5-2.0 keV and 2-10 keV
fluxes (averaged from the three detectors which differed from the mean
by less than  $\sim 25\%$) were $5.1 \times 10^{-14}$~\ergcms and
$8.7 \times 10^{-14}$~\ergcms.  If this
source is associated with NGC 3998 then the implied luminosity is $2
\times 10^{39}$\ergs, making it an ``ultra-luminous X-ray source''
(ULX).  A multi-color disk (MCD) black body model resulted in an
unacceptable fit, with $N_H = 0. \rm \ cm^{-2}$, $kT_{in} = 0.7$ keV,
and $\chi^2$/dof = 53.1/32.  
The observed X-ray spectrum is consistent with ULXs,
often observed with power-law spectra \citep[e.g.,][]{st01, fo02}.
However of course a power-law spectrum with $\Gamma \sim 1.9$ is also
consistent with a background AGN.
The {\it XMM-Newton} 2-10 keV logN-logS shown in
\citet{ha01} implies that $\sim 10$ background sources $\rm deg^{-2}$ would be
expected at this flux level or higher.  X-1 is $1.6'$ from the nucleus
NGC 3998, and so only $\sim 0.02$ background sources are expected at
this flux or higher within this area, implying that it is unlikely
that this is a background source. 
Since we used a source-free region for the background spectra to
avoid inaccurately subtracting flux due to NGC 3998 (as would result
from using a ``local'' background), however the
spectra of the ULX spectra then contain some contamination from NGC 3998.
To quantify this, we used our SciSim 
simulation of the EPIC data and estimate that 0.19\% of the flux of
NGC 3998 is contaminating the X-1 source regions, or $\sim 2 \times 
10^{-14}$\ergcms = 24\% of the observed flux of X-1.

%\clearpage
 
%\clearpage

We also show in Figure \ref{pn_x-1_fig} the U band image.  After
adjusting the astrometry by applying a $\sim 1''$ shift (derived by
aligning the FIRST position of the nucleus of NGC 3998 and two USNO-B
stars), an additional USNO-B source, 1454-0216445, is coincident with
a U band source and NGC 3998 X-1.  The U band source is detected with a
signal-to-noise ratio of 9.3 and a magnitude $19.9 \pm 0.3$ (derived using the
tool ``omsource''; note that here the statistical error on the U
magnitude is larger than the calibration
uncertainties).  The USNO-B 2nd epoch (1974.9) R and B magnitudes 
are 20.0 and 21.6.  The USNO-B 1st epoch values are brighter by
0.5 and 0.7 magnitudes, respectively. However the uncertainty of
USNO-B magnitudes is typically $\sim 0.3$ \citep{mo03}, and since
these magnitudes are 
close to the flux limits of the survey (V $\sim 21$), their uncertainties are
probably somewhat larger, particularly in the case of the B band
magnitude.  Therefore the difference between the two 
USNO-B epochs is probably just due to statistical fluctuation.
Similarly the unusual ``concave'' spectrum implied by the USNO-B
second epoch R and B magnitudes and the OM U band magnitude is
largely due to the questionable B band magnitude.

\section{Discussion}
As discussed in \citet{pe00}, the X-ray spectrum of NGC 3998 is fit well in the
0.1-100 keV bandpass with a simple power-law model that is only
moderately absorbed ($N_H \sim 3 \times 10^{20} \rm \ cm^{-2}$ in
contrast to the Galactic value of $\sim 1 \times 10^{20} \rm \
cm^{-2}$).   
%This absorption is likely intrisic to NGC 3998 but
%extranuclear.  
We do not find any evidence for large-scale (i.e.,
$\Delta L/L > 25\%$) variability on short ($< 1$ day)
time scales but NGC 3998 may have varied in the 0.5-2.0 keV bandpass
on long time scales.  Our main goal is to assess the 
physics of the accretion in NGC 3998, and below we compare these observed X-ray
properties of NGC 3998 with those observed in other AGN, and compare
the SED of NGC 3998 with Seyfert 1 and
RIAF models. 
We also speculate on the constraints that our results can place on
the geometry of the nucleus of NGC 3998.

\subsection{Comparison with Seyferts}
The presence of broad $H\alpha$ in NGC 3998 suggests that the nuclear
region might be 
similar in structure to ``typical'' Seyfert 1 galaxies.  
The X-ray slope of NGC 3998 ($\Gamma = 1.9$) is consistent with
typical Seyfert 1 galaxies, and the 2-10 keV luminosity of NGC 3998
($2.6 \times 10^{41}$~\ergs for $F_{\rm 2-10 \ keV} = 1.1 \times
10^{-11}$\ergcms) is only marginally low for
typical Seyfert galaxies \citep{na97a}.   However, NGC 3998 differs in
two important respects from typical Seyfert 1 galaxies.  First, the
lack of strong X-ray variability is in contrast with typical 
Seyferts, particularly on short time scales, since the trend there is
for variability to increase with decreasing 
luminosity \citep{na97a}.  \citet{pt98} suggested that this is typical
of LINERs and LLAGN (including NGC 3998) and that the
X-ray source regions are larger in LINERs and LLAGN than in typical
Seyferts.  One possibility is that the X-ray production is occurring in
optically-thin flows (see below) rather than flares in
optically-thick accretion disks.
%An alternative possibility is that LINERs
%tend to have larger black hole masses than Seyfert galaxies, although
%the black hole masses required are not  \citep{aw00}. 

\begin{figure}
%\plotone{n3998_sy1_fek.eps}
\plotone{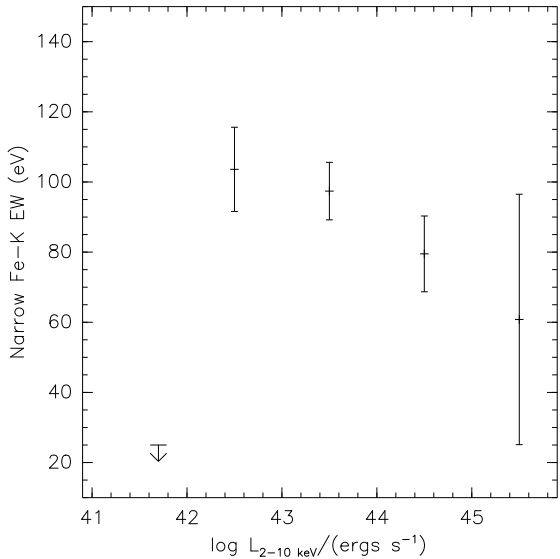}
\caption{Observed narrow-line Fe-K EWs as a function of
  luminosity for Seyfert 1 galaxies from \citet{na97c} with the
  upper-limit for NGC 3998 also shown. \label{ewfig}}
\end{figure}

The X-ray spectrum is also remarkably featureless in contrast with
typical Seyfert 1 galaxies, particularly with the lack of any Fe-K
emission at 6.4 keV.  The upper-limit to the equivalent width (EW) in
the {\it XMM-Newton} and {\it BeppoSAX} joint fit is 25 eV.  Figure~\ref{ewfig}
shows the distribution of narrow-line EWs measured in a sample of
Seyfert 1 galaxies \citep{na97c}, where we also show the upper-limit
for NGC 3998.  The trend in Seyfert 1s is to have higher EW Fe-K lines
with decreasing luminosity, and accordingly the lack of a line in NGC
3998 is conspicuous.  A cautionary point is that Figure~\ref{ewfig} is
based on Seyfert 1s with Fe-K lines and does not account for Seyfert
1s lacking Fe-K lines.  However, the vast majority of (X-ray bright)
Seyfert 1 galaxies appear to exhibit some level of Fe-K emission
\citep{na97b}, with very few Seyfert 1s being known that lack a
Fe-K line when the spectra are of sufficient quality for Fe-K emission
to be detected (T. Yaqoob, priv. comm.). Thus, it is likely that the
accretion physics and/or 
geometry in NGC 3998 differs somewhat from typical Seyfert 1 galaxies.

%\clearpage

%\clearpage

\subsection{Radiatively-Inefficient Accretion Flows}
By integrating the best-fitting power-law to the {\it XMM-Newton} and
{\it BeppoSAX} data ($\Gamma = 1.88$ with normalization $3.8 \times 10^{-3}
\rm \ photons \ keV^{-1} \ cm^{-2} \ s^{-1}$) we calculate the
unabsorbed 0.1-100 flux to be $6.4 \times 10^{-11}$\ergcms, or a
luminosity of $\sim 1.5 \times 10^{42}$ \ergs. Using the bulge
velocity dispersion of 304 km s$^{-1}$ from \citet{mc95} and the bulge
dispersion/black hole mass relation given in \citet{tr02} we estimate
the black hole mass of NGC 3998 to be $\sim 7 \times 10^{8}
M_{\odot}$.  This implies an Eddington luminosity ($L_{Edd} = 1.25 \times
10^{38}\frac{M}{M_{\odot}}$) of $9 \times 10^{46}$ \ergs, so  
$L/L_{Edd} = 2 \times 10^{-5}$ (note that we get a similar result if
we use the typical LLAGN ratio of $L_{2-10 \rm \ keV}/L_{bol} = 0.15$ found
by Ho [1999]).
%Assuming an accretion power efficiency of 10\%
%results in an accretion rate of $\dot{m} = 3 \times 10^{-4}$ in
%Eddington units. 
One possibility is that the low luminosities in NGC 3998 are
reflecting low accretion rates. 
Recent accretion modeling suggests that at low accretion rates
($\lesssim$ a few percent of the Eddington rate; see Narayan,
Mahadevan, \& Quataert 1998) 
 the
flow will have low particle densities, and in turn may be optically thin
and radiatively inefficient.  In general these models predict that the X-ray
spectrum from an RIAF is dominated by inverse-Compton scattering (of
synchrotron emission) and results in somewhat steep X-ray slopes (as
opposed to the flat slopes expected from bremsstrahlung emission,
e.g., see Mahadevan 1997).  For example, in the case of
convection-dominated accretion flows (CDAFs), X-ray slopes of $\sim 2$
are expected when  
$L_X/L_{Edd} > 10^{-7}$ \citep{ba01}. 
The X-ray temporal and
spectral properties of NGC 3998 described above are similar to other
LINERs thought to contain RIAFs \citep[e.g.,][]{pt98, qu99b} (although
again the X-ray spectral slope is consistent with typical Seyfert 1s also). 

\subsection{Spectral Energy Distribution}
We turn now to comparison of the SED of
NGC 3998 to the predictions of various models.  The data used to
derived the SED of NGC 3998 is listed in Table \ref{sedtab} and is shown
in Figure~\ref{sedplot}.  The X-ray data shown has been ``unfolded'' to remove
the spectral response of the detectors and
%Only the {\it XMM-Newton}, {\it ASCA} SIS0 and
%{\it BeppoSAX} LECS, MECS3 and PDS data are shown for clarity (note also
%that t
have been binned more so than in Figure \ref{allplfig} for display
purposes.
The points marked with open circles can be considered to be upper-limits
since they are sampling 
radii $> 10$\arcsec.  In the near-IR the large aperture points are clearly 
dominated by galactic emission (and accordingly the small aperture
points contain a large contribution from galactic light).
However in the radio to mid-IR region the
radiation appears to be fairly compact (i.e., unresolved to aperture sizes
less than $\sim 10\arcsec$, see also Knapp et al. 1996).  Also variability
is evident among several of the radio and (possibly also) mid-IR measurements
listed in Table \ref{sedtab}, which also implies that the emission is
compact.  Note that SEDs for NGC 3998 were also presented in \citet{fa94} and
\citet{pe00}, however those papers assume the HST FOC UV flux, which is a
a factor of $\sim 2-4$ higher than our OM measurement.  This higher
UV flux results in a SED that resembles a typical AGN, while our lower UV
flux results in a SED more consistent with the sample of low-luminosity AGN
discussed in \citet{ho99}.  It is also evident that in the optical
band, only the HST data with $\sim 0.1''$ apertures are sufficient to
exclude large contributions of extra-nuclear flux.  Given the large
low-mass star contribution to near-IR emission of early-type galaxies,
this is also true for the near-IR band.  The relative amounts of flux
in the OM U band and HST FOC U band suggests that this contribution is
at least 90\%, and likely to be much larger at longer wavelengths.

We assessed the impact of extinction using the extinction equations in
\citet{ca89}.   \citet{kn85} found that the upper-limit for HI
absorption to the nucleus of NGC 3998 was $\sim 4 \times 10^{20} \rm \
cm^{-2}$.  Including the Galactic contribution of $\sim 1 \times
10^{20} \rm  \ cm^{-2}$ results in a conservative estimate of $N_H
\sim 5 \times 10^{20} \rm \ cm^{-2}$.  This value is consistent with
the X-ray observations here with the exception of the {\it ASCA} data,
however again we note that the calibration of the low energy response
of the {\it ASCA} CCD detectors is somewhat uncertain.  This corresponds to
$A_V \sim 0.22$ \citep{je74}, and from \citet{ca89} we derive that
$\sim$ 50\%, 30\% and 25\% of the flux at UVW2, U and B, respectively,
would be absorbed.  This level of extinction does not impact our conclusions.

%\input{sed_table}

%\clearpage

\begin{deluxetable*}{llll}
%\rotate
\tablecaption{Luminosity Data for NGC 3998\label{sedtab}}
\tabletypesize{\scriptsize}
\tablehead{
\colhead{$\nu$ (Hz)} & \colhead{$\nu L_\nu \ (ergs \ \rm s^{-1})$} &
\colhead{Aperture (\arcsec)} & \colhead{Reference}
}
\startdata
$3.2 \times 10^{8}$ & $6.5 \times 10^{36}$ & 60 & \citet{re97}\\
$6.0 \times 10^{8}$ & $1.4 \times 10^{37}$ & 30 & \citet{hu84}\\
$1.4 \times 10^{9}$ & $3.3 \times 10^{37}$ & 24 & \citet{hu80}\\
$1.4 \times 10^{9}$ & $3.4 \times 10^{37}$ & 45 & \citet{co98}\\
$1.4 \times 10^{9}$ & $3.1 \times 10^{37}$ & 5 & FIRST\\
$2.7 \times 10^{9}$ & $7.0 \times 10^{37}$ & 6 & \citet{co78}\\
$5.0 \times 10^{9}$ & $9.1 \times 10^{37}$ & 0.5 & \citet{hu84}\\
$5.0 \times 10^{9}$ & $1.0 \times 10^{38}$ & 0.008 & \citet{hu82}\\
$5.0 \times 10^{9}$ & $9.6 \times 10^{37}$ & 0.007 & \citet{fi02}\\
$8.1 \times 10^{9}$ & $1.8 \times 10^{38}$ & 6 & \citet{co78}\\
$1.5 \times 10^{10}$ & $2.1 \times 10^{38}$ & 0.1 & \citet{hu84}\\
$3.0 \times 10^{12}$ & $8.6 \times 10^{41}$ & 180 & IRAS\\
$5.0 \times 10^{12}$ & $6.6 \times 10^{41}$ & 90 & IRAS\\
$1.2 \times 10^{13}$ & $3.7 \times 10^{41}$ & 45 & IRAS\\
$1.5 \times 10^{13}$ & $1.0 \times 10^{42}$ & 8 & \citet{wi85}\\
$2.5 \times 10^{13}$ & $8.4 \times 10^{41}$ & 45 & IRAS\\
$3.0 \times 10^{13}$ & $2.1 \times 10^{41}$ & 8 & \citet{wi85}\\
$3.0 \times 10^{13}$ & $3.8 \times 10^{41}$ & 6 & \citet{wi85}\\
$3.0 \times 10^{13}$ & $3.7 \times 10^{41}$ & 6 & \citet{wi85}\\
$8.6 \times 10^{13}$ & $1.5 \times 10^{42}$ & 5 & \citet{wi85}\\
$1.4 \times 10^{14}$ & $5.5 \times 10^{42}$ & 3 & \citet{al00}\\
$1.4 \times 10^{14}$ & $4.0 \times 10^{42}$ & 5 & \citet{wi85}\\
$1.4 \times 10^{14}$ & $5.1 \times 10^{42}$ & 11 & \citet{lo82}\\
$1.4 \times 10^{14}$ & $8.3 \times 10^{42}$ & 15 & \citet{fr78}\\
$1.4 \times 10^{14}$ & $4.0 \times 10^{42}$ & 4 & 2MASS Point Source Catalog\\
$1.8 \times 10^{14}$ & $6.8 \times 10^{42}$ & 5 & \citet{wi85}\\
$1.8 \times 10^{14}$ & $9.0 \times 10^{42}$ & 11 & \citet{lo82}\\
$1.8 \times 10^{14}$ & $1.4 \times 10^{43}$ & 15 & \citet{fr78}\\
$1.8 \times 10^{14}$ & $6.4 \times 10^{42}$ & 4 & 2MASS Point Source Catalog\\
$2.5 \times 10^{14}$ & $9.0 \times 10^{42}$ & 3 & \citet{al00}\\
$2.5 \times 10^{14}$ & $7.8 \times 10^{42}$ & 5 & \citet{wi85}\\
$2.5 \times 10^{14}$ & $9.9 \times 10^{42}$ & 11 & \citet{lo82}\\
$2.5 \times 10^{14}$ & $1.5 \times 10^{43}$ & 15 & \citet{fr78}\\
$2.5 \times 10^{14}$ & $7.4 \times 10^{42}$ & 4 & 2MASS Point Source Catalog\\
$3.8 \times 10^{14}$ & $1.3 \times 10^{41}$ & 0.1 & This work\\
$5.5 \times 10^{14}$ & $1.0 \times 10^{41}$ & 0.1 & This work\\
$5.9 \times 10^{14}$ & $1.6 \times 10^{41}$ & 0.1 & \citet{fa94}\\
$8.7 \times 10^{14}$ & $2.5 \times 10^{41}$ & 3 & This work\\
$8.8 \times 10^{14}$ & $>$$5.5 \times 10^{40}$ & 0.1 & \citet{fa94}\\
$1.4 \times 10^{15}$ & $1.8 \times 10^{41}$ & 2 & This work\\
$1.8 \times 10^{15}$ & $>$$4.1 \times 10^{41}$ & 0.1 & \citet{fa94}\\

\enddata
\end{deluxetable*} 

%\clearpage

\subsubsection{Comparison with RIAF Models}

For accretion onto a black hole, the gravitational in-fall timescale
can be shorter than the radiative timescale; the accretion energy is
then stored as thermal energy instead of being radiated (e.g., Rees et
al. 1982; Narayan \& Yi 1994).  
In the past few years, analytical and
numerical work has shown that the physics of such RIAF models is far
more complex than originally anticipated.  In particular, very
little mass available at large radii actually accretes into the
black hole; most of it is lost to an outflow (i.e., in an adiabatic
inflow/outflow solution, or ``ADIOS'') or circulates in convective
motions \citep[e.g.,][]{bl99,st99,ig00,qu00,ha02,ba02}. The accretion
of slightly rotating gas also may result in a substantially reduced
accretion rate relative to the Bondi rate\citep{pr03}.
%(e.g., Blandford \& Begelman 1999; Stone et al. 1999;
%Igumenshchev \& Abramowicz 1999; 2000; Quataert \& Gruzinov 2000;
%Hawley \& Balbus 2002).  
Following a proposal due to Blandford \&
Begelman (1999), we can parameterize the radial variation of the gas
density in the flow with a parameter $p$, where $\rho \propto r^{-3/2
+ p}$ (equivalently, the accoretion rate varies with radius as $\dot m
\propto r^p$); $p \approx 1/2-1$, rather than $p = 0$ (as in the
original analytical advection-dominated accretion flow [ADAF] models
of Narayan \& Yi 1994), is favored by the simulations, implying much
lower gas densities close to the black hole.
%%
%%(although instabilities may arise in
%%advection-dominated accretion flows (ADAFs) resulting in variability
%%on time scales shorter than $\sim 
%%R_g/c$, or $\sim 7$ ks in the case of NGC 3998). 

\begin{figure*}
%\plottwo{n3998_sed_a_29Dec03.eps}{n3998_sed_b_29Dec03.eps}
\plottwo{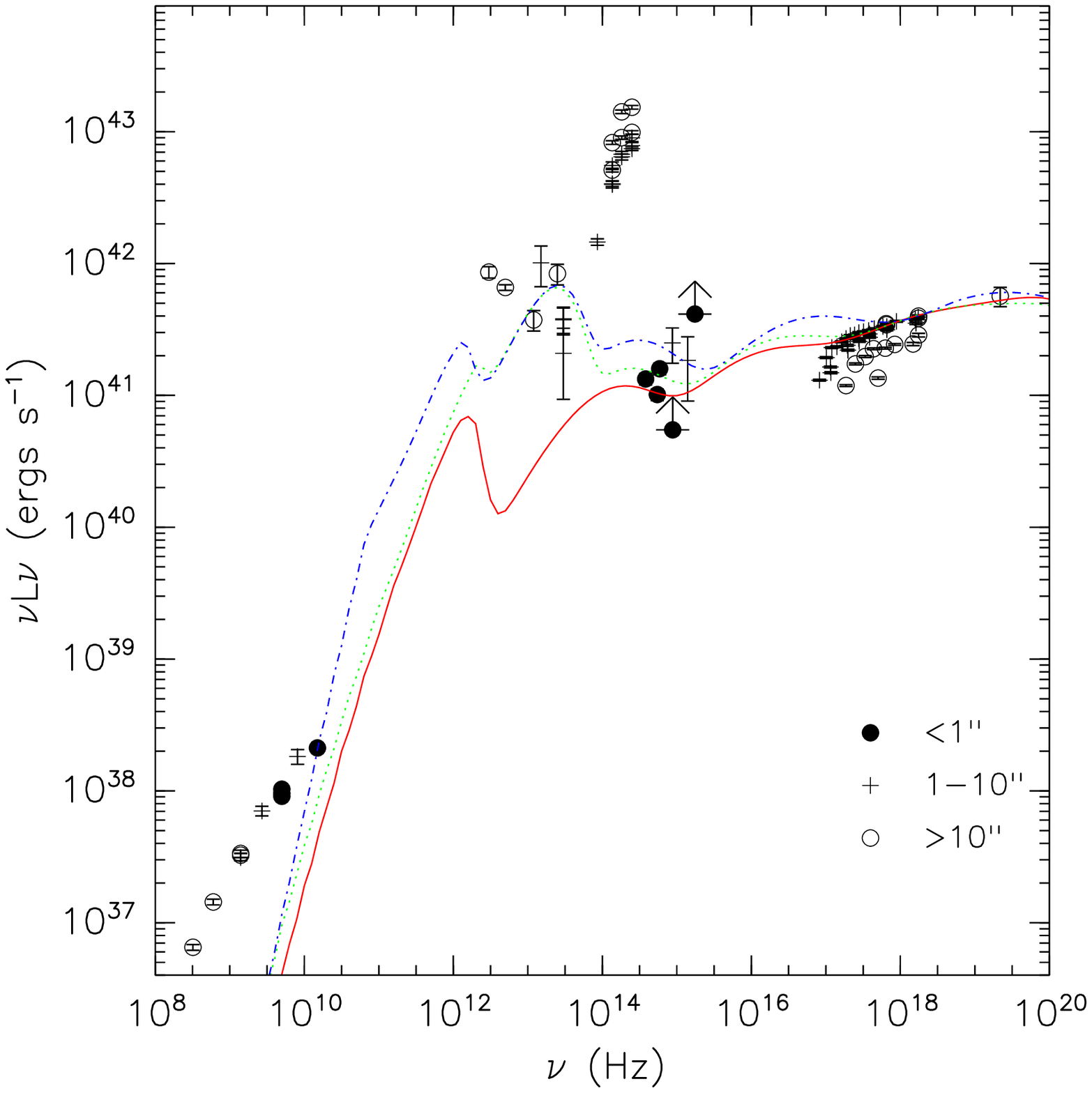}{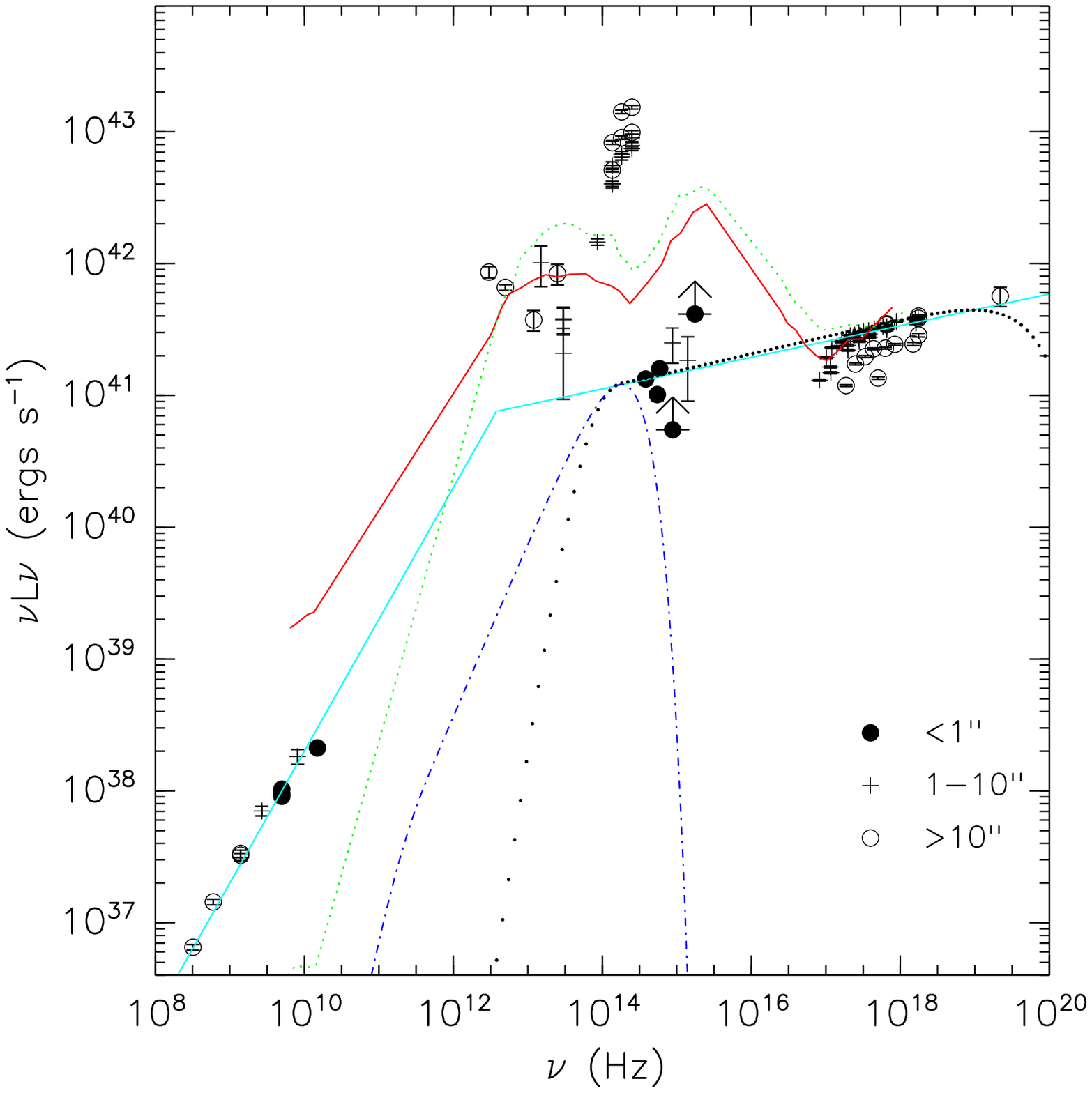}
%\plottwo{f8a_bw.eps}{f8b_bw.eps}
\caption{The SED of NGC 3998 plotted with (left panel) two RIAF models
  and (right panel) the radio-load (red solid line) and radio-quiet
  (green dotted line) mean QSO SEDs from
  \citet{el94}.  The U and UV points
  (at $\nu \sim 10^{15}$) are due to the OM, and the X-ray points are
  due to the spectral fits discussed in the text (note that the data
  points have been 
  ``unfolded'' to account for the spectral responses of the detectors
  and hence are somewhat model dependent).  The turn-over in the X-ray
  data at low energies is due to absorption (with $N_H \sim 3 \times
  10^{20} \rm \ cm^{-2}$).
  %, and the dashed line
  %shows the extrapolation of the X-ray power-law model (i.e, with an energy
  %index of 0.9).  
  The X-ray data points with lower luminosities are
  due to {\it ASCA} (see text).  The RIAF models
  are computed with an inner ADAF (green dotted line; $\dot{m}=3 \times
  10^{-3}$ in Eddington units,
  $\delta=0.01$) and an inner ADIOS (blue dot-dashed line; $\dot{m}=3
  \times 10^{-3}, \delta=0.3, p=0.4$), both with an outer thin disk with a
  transition radius of 300 $R_{s}$ and outer radius of $10^5 R_{s}$.  
  We also plot 
  the ADAF model with no outer thin disk for comparison (red solid
  line). The right panel also shows a Comptonized disk spectrum
  (black bold dotted line).  The Comptonization
  model used is based on a Wien spectrum approximation, and
  we show the 
  equivalent multi-color disk model for comparison (blue dot-dashed
  line; 
  %the parameters were $\dot{m} = 3.6 \times 10^{-6}$, and
  %inclination angle = 55$\degr$).
  the parameters were $\dot{m} = 4.1 \times 10^{-6}$, and
  inclination angle = 60$\degr$).  A simple broken power-law model
  with a radio slope of 0.0 and an IR-to-Xray slope of 0.9 is also shown
  in the right panel as a representative synchrotron model.
\label{sedplot}}
\end{figure*}

%Quataert \& Narayan (1999) showed that there are significant
%degeneracies in comparing theoretical RIAF spectra to observed data.
Figure \ref{sedplot}a shows several RIAF models for the SED of NGC 3998.
\citet{qu99a} showed that there are significant degeneracies in comparing
theoretical RIAF spectra to observed data so these results should be 
interpreted as representative models, not unique ``fits.''  The solid line in
\ref{sedplot}a is an ADAF with an
accretion rate of $3 \times 10^{-3}$ (hereafter accretion rates are
given in Eddington units) and $\delta = 0.01$ (= fraction of the
accretion energy transfered to electrons).  The dotted line is the ADAF model
with an outer thin disk starting at a transition radius $r_t = 300
R_{s}$ ($R_s$ = the Schwarzschild radius).  
The dot-dashed line is an
ADIOS model with 
$\delta = 0.3$ and $p = 0.4$ and a similar accretion rate at the outer
edge of the flow, also with an outer thin disk (the so-called
multi-color disk (MCD) model) with $r_t = 300 R_s$.  
Note that ADIOS models result in similar spectral models to CDAF
models. The thin 
disk (for both models) has an 
outer radius of $10^5 R_s$, an inclination of 60 degrees, and $\dot{m}
\approx 3 \times 10^{-3}$ (hereafter all $\dot{m}$ values are quoted
in Eddington units).  
Given theoretical uncertainties about electron heating, values of
$\delta << 1$ to $\delta  \sim 1$ are
plausible (e.g., Quataert \& Gruzinov 1999), and the two models chosen in fact
span this range.
These models fit the IR-X-ray emission
reasonably well, except of course for the large contribution of
stellar flux at near-IR wavelengths.  The (nuclear) IR emission is
dominated by the (putative) thin disk 
component while the UV-X-rays are dominated by the RIAF, and in fact
$r_t$ and the accretion rate at $r_t$ were chosen to approximately
match the IR flux.  The smaller $r_t$ value could have been chosen
which would have resulted in a larger thin-disk flux, and
correspondingly a smaller $\dot{m}$ would have been chosen.  Therefore
the values of $\dot{m}$ and $r_t$ are only constrained to within
$\sim$ an order of magnitude.
The X-ray emission in these RIAF models is dominated by
inverse-Compton scattering which results in approximately in a
power-law continuum. 
The ADIOS model has a lower
$\dot{m}$ and density near the black hole, but a larger temperature.  In
that situation, the  Compton peaks are more 'separated' (since the
energy gain per scattering is proportional to temperature) and the
peaks are more distinct.  This is a
generic difference between high $\dot{m}$ ADAF type models and lower $\dot{m}$
models.  Since the X-ray spectrum of NGC 3998 is fairly featureless
lower temperature models are more consistent with the X-ray data.
Note that in general different thermal line emission characteristics are
predicted for these 
different types of models, but these lines are not detectable at this
$\dot{m}$ \citep{perna00}. 
Neither
model accounts for the radio emission, in spite of the fairly large
outer disk radius of $10^5 R_s$.  The most restrictive constraint in this
regard is the VLBI 5 Ghz point which has a beam size of only 7 mas, or
$\sim 0.7$ pc, and is clearly nuclear in origin. 
%The dashed line shows the
%best-fitting power-law extrapolated to $\nu=10^{14}$ Hz.

%This can be remedied by
%either reducing the accretion rate in the model or increasing the transition
%radius.  

%\clearpage

%\clearpage

\subsubsection{AGN Spectral Models}
In Figure \ref{sedplot}b we also plot the median SEDs for radio-load and radio-quiet AGN from
\citet{el94}, rescaled to the flux of NGC 3998 at $\nu =
10^{18}$ Hz.  The QSO SEDs grossly overestimate the blue-UV emission,
i.e., NGC 3998 is lacking the ``big blue bump'' that is considered to
be due to an optically-thick
accretion disk.  
However, this is not unexpected since the accretion
rate is much lower than that of a typical QSO, and MCD temperatures scale
as $\dot{m}^{1/4}$ \citep{fr02}.  
We also show an approximate
Comptonized thin-disk model that fits the X-ray data from NGC 3998
without over-predicting optical or IR flux.  This thermal
Comptonization model (implemented in XSPEC as ``compTT''; see
Titarchuk 1994, Hua \& Titarchuk 1995) uses the
Wien approximation to black body flux for computational efficiency,
however this should be a reasonable approximation.  In this model
there are 5 parameters: the overall normalization, the temperature of the Wien
component, the temperature and optical depth ($\tau$) of the hot
plasma, and a geometry switch to select a spherical or disk geometry.   
In general inverse-Compton spectra tend to be power-laws with a slope
of $-\frac{\ln \tau}{\ln A}$, where $A$ is the mean amplification of
the photon energy per scattering \citep{ry79}.  Therefore it is not
surprising that for a given Wien and plasma temperature combination,
$\tau$ could be adjusted to result in our observed power-law slope and
hence give a good fit.  The fits generally resulted in a power-law
extrapolated from the peak of the Wien component, and we therefore
initially fixed the 
temperature of the Wien spectrum at $10^{-4}$ keV ($\sim 10^{3} K$) to
see if the power-law could be extended to the optical bandpass.
We tried hot plasma temperatures of 100 and 200 keV \citep[see][]{pi95},
and fitted
for the model normalization and $\tau$ as free
parameters.  The best-fit optical depths were 0.3 and 0.1 for a disk
geometry and plasma temperatures of 100 and 200 keV, respectively.
With a spherical geometry the best-fit optical depths were 0.5 and 1.0
for plasma temperatures of 100 and 200 keV.  Clearly the value of
$\tau$ is not meaningful since our data do not constrain the geometry
or temperature of the hot plasma.  These fits
(simultaneously to the {\it XMM-Newton} and {\it BeppoSAX} data)  resulted
in $\chi^2$/dof = 1653-1654/1567, or basically equivalent to the
pure power-law fit.  
The fit with a plasma temperature of 200 keV and a disk geometry is
shown in Figure \ref{sedplot}.  We 
computed a MCD spectrum (with inner 
radius = $3 R_s$ and outer radius = $10^5 R_s$)
consistent with the Wien component by assuming an inclination angle of
60 degrees and adjusting $\dot{m}$, which resulted in $\dot{m} = 4.1
\times 10^{-6}$. 
Such a model accounts for the spectrum of
NGC 3998 from the optical to $\sim 100$ keV, i.e., the optical flux is
consistent with the extrapolation of the power-law observed in the
X-ray band, particularly if we allow for variability in the FOC F480LP
flux (the offset
between the PDS point in Figure \ref{sedplot} and the Comptonized
spectrum model is due in part to the
fact that the X-ray data were unfolded assuming the simple
power-law spectrum extending to 100 keV).  However this model
requires that a 
{\it negligible} fraction of the emission longward of the optical is due to
the AGN in NGC 3998.  This may be unlikely given the compactness of
the radio and mid-IR emission.

Of course there is some degeneracy between the
inclination angle and accretion rate of the disk in this scenario,
with an angle of 0$\degr$ and 
$\dot{m} = 2.0\times 10^{-6}$ also being consistent with the Wien
component.  
However the luminosity of the MCD disk component is not
arbitrary.  The MCD spectrum is given by the sum of blackbody spectra with
temperatures varying as $r^{-3/4}$ where $r$ is radius of each point
on the surface of the disk \citep{fr02}.  This results in a spectrum
similar to a single blackbody emitting at the peak temperature of the
disk, and for given black hole mass the peak temperature scales as
$\dot{m}^{1/4}$.  Of course the luminosity of a blackbody emitting
over a given surface area is proportional to $\nu^{4}$, where h$\nu
\sim 3$kT.   
%, and additionally fixing the
%normalization to the  
%luminosity expected from the MCD component at temperatures
%differing from $10^{-4}$ keV by more than $\sim 1 dex$ resulted in
%power-law slopes that were not consistent with the X-ray spectra.
%This is because once the Wien component normalization is fixed, the
%only free parameter that can be adjusted to result in the observed
%X-ray fluxes is the optical depth of the hot plasma, and the power-law
%slope due to Comptonization is a function of the optical depth.  
Therefore we could
not match the 
luminosity of the MCD component for other disk temperatures by
simply adjusting the accretion rate since the expected luminosity of
the Wien component would move in the $\log L_\nu - \nu$ plane along a slope
of 4 rather than 0.9.  This makes the Comptonized MCD model
inconsistent with the observed X-ray power-law if the peak disk temperature
is below $\sim 10^{-4}$ keV (i.e., the Wien component normalization
would be larger than the MCD expectation), and also at higher temperatures this
model is only 
consistent with the data if the MCD component is attenuated (e.g., by
inclination effects). Also note that the optical (UV) flux would not be
due to the accretion disk for disk temperatures in
excess of $\sim 10^{-4} (10^{-3})$ keV. Conversely, assuming that the
optical and UV data are due to a thin disk requires that $\dot{m} \la
10^{-5}$.  Finally, note that a standard $\alpha$ disk is both stable
and optically thick when $\dot{m}$ is in the range discussed here
($10^{-6} < \dot{m} < 10^{-5}$). 

%This is due in part
%to the relatively high mass of the black hole in NGC 3998, which pushes most
%disk emission into the IR.  
\subsubsection{Jet Models}
In addition to a very low accretion rate, another possibility is that both
the radio and X-ray flux from NGC 3998 are due to a jet, as has been
suggested recently for LLAGN such as NGC 4258 \citep{yu02}, IC 1459
\citep{fa03} and NGC 4594 \citep{pe03}. Along these lines we show a
simple broken power-law 
model in Figure \ref{sedplot}, where we found that the radio data
were consistent with a slope of $\sim 0$ (i.e., with $\nu L_{\nu}
\propto \nu^{1}$) and break frequency of $\sim 10^{12}$ Hz.  This is 
obviously an overly simplistic model but here the radio slope is due
to optically-thick synchrotron emission and the X-ray emission is due
to optically-thin synchrotron emission, modeled by \citet{ma01} as
being due to the outer, ``post-shock'' portion of the jet. The
Markoff-Falcke model also predicts a mid-IR ``bump'' due to
synchrotron emission from the inner, ``pre-shock'' jet which may be
consistent with the mid-IR fluxes observed in NGC 3998. Alternatively,
the synchrotron emission may break to a flatter 
slope and the X-ray is again also due to inverse-Compton emission, and
the mid-IR and/or optical-UV flux is due to a standard $\alpha$ disk
\citep{ma03}. 
Clearly these models would also be consistent with the SED of NGC
3998 and may imply accretion rates possibly orders-of-magnitude higher
than the RIAF and Comptonized $\alpha$-disk models since a large
fraction of the accretion energy would be released kinetically in the
jet. However, as was the case in the SED modeling discussed above, any
$\alpha$ disk in this scenario would likely 
still need to be truncated to avoid over-predicting the optical and UV
fluxes.  Also, the lack of any short-term, large-amplitude X-ray
variability may argue against the X-ray emission being dominated by a 
(relativistic) jet.

\subsection{Constraints on the Nuclear Geometry of NGC 3998}
The Fe-K lines in Seyfert 1 galaxies are assumed to originate from
reflection from a thin disk and/or matter further out such as the
putative ``torus'' in the standard model \citep{ya02}.  If the Fe-K emission
is dominated by a thin disk, then the maximal solid angle as seen from
the X-ray source would be $\sim 2\pi$, and the solid angle of a
toroidal geometry of matter most likely would not exceed this.  If we
assume that the EW $\sim 100$ eV Fe-K line typically observed in
Seyfert 1 galaxies is produced by a reflection from optically-thick
material with $\Omega \sim 2\pi$, then our upper-limit of $\sim 25$ eV
implies that $\Omega/2\pi \sim 0.25$ in NGC 3998.  Interestingly,
\citet{pe00} report a similar result of $\Omega/2\pi <0.21$ for either
a neutral or ionized Compton reflection hump based on reflection model
fits to the {\it BeppoSAX} spectrum of NGC 3998.  We also computed an
upper-limit for a disk line model at 6.4 keV and obtained $\sim 30$
eV.  A $L_{2-10 \rm \ keV} = 10^{42}$ \ergs~ Seyfert 1 would typically
exhibit a EW = 300 eV disk line, which implies $\Omega/2\pi < 0.1$.
This suggests that the solid angle of {\it any} optically-thick
material in NGC 3998 is less than at most 25\% of that typically
observed in Seyfert 1 galaxies.  

%However, for a thin-disk geometry this is not restrictive on the size
%of the thin disk itself since the size and position of the X-ray
%source irradiating the disk is arbitrary.

We can roughly constrain the size of an outer thin disk using the line
information.  Assuming a disk geometry with the X-ray source at a
height $h$ above the center of the disk with inner and outer radii
$r_{in}$ and $r_{out}$, the solid angle subtended by the disk is given
by $\Omega = 2\pi[(1 + (\frac{r_{in}}{h})^2)^{-1/2} - (1 +
(\frac{r_{out}}{h})^{2})^{-1/2}]$.  The X-rays are produced in the
inner part of the RIAF, implying $h \sim 10-30 R_s$.  It is therefore
likely that $r_t >> h$.  This implies that $\frac{\Omega}{2\pi} \sim
h(\frac{1}{r_{in}} - \frac{1}{r_{out}})$ and therefore
$r_t \ga \frac{h}{\frac{\Omega}{2\pi}}$, or $r_t \ga 100-300 R_s$ for
$\frac{\Omega}{2\pi} \la 0.1$.  This is consistent with Figure \ref{sedplot}
where we 
assumed $r_{in} = r_t \sim 300 R_s$.  A possible caveat to this analysis is 
that NGC 3998 may contain an optically-thick disk with $r_{in} = 3
R_s$ but whose  
surface layers are fully ionized.  This could result in little or no Fe-K
emission \citep{do01, ba02}. 

%For the preferred
%transition radius of $r_t \sim 300$ from Figure $\ref{sedplot}$ we
%have the hierarchy $h \ll r_{in} \ll r_{out}$ and the solid angle
%subtended by the disk reduces to $\Omega/2 \pi \approx h/r_t \sim
%0.1$, consistent with the lack of a detectable iron line.

\subsection{The Nature of the ULX NGC 3998 X-1}
If this is source is associated with NGC 3998, then the optical
luminosity of the source ($M_R \sim$ -11) implies several to tens of
supergiant stars or a cluster of hundreds to thousands of low-mass
stars, if the optical flux is stellar.   
If the optical variability observed between the two USNO-B epochs
(separated by of order 25 years) were real it would argue against
this, although again the variability is probably not statistically
significant.   
In addition,
the high U-band absolute magnitude ($M_U = -10.9$) and very blue
color imply that any such cluster would be composed of
early-type stars.  This rules out a globular cluster with late-type
stars which would be expected in a early type galaxy.  
%The second epoch USNO-B B magnitude corresponds to
%a flux of $\sim 1 \times 10^{-28} \rm \ ergs \ cm^{-2} \ s^{-1} \
%Hz^{-1}$, or $\nu F\nu \sim 7 \times 10^{-14}$\ergcms.  
Using $\log F_X/F_{opt}$ = log($F_{0.5-2.0 \ \rm keV}$) + R/2.5 + 5.7
\citep{ho03, sz04, no04}, we derive $\log F_X/F_{opt} = 0.4$ for the ULX
which is typical of AGN found in
the CDF-S \citep{sz04, no04} and is much larger than values typical of
stars ($\log F_X/F_{opt} << -1$) and somewhat smaller than values typical
of low-mass X-ray binaries ($\log F_X/F_{opt} > 1$;  Bradt \&
McClintock 1983; although realistically we would be
observing an X-ray binary in a cluster). 
%This is larger than the typical
%$F_X/F_{opt}$ observed in QSOs \citep[$\sim 0.1$][]{el94}.  
The most likely identification for this source is therefore a
background AGN, although as stated in \S\ref{ulxsec}, the detection of
a background source at this flux was unlikely..

\section{Summary}
We have determined the tightest constraint to date on Fe-K emission in
NGC 3998, specifically EW $< 25$ eV for a narrow line at 6.4 keV.  Our
analysis of {\it XMM-Newton} data combined with archival {\it ASCA} and {\it BeppoSAX}
data shows that the spectrum is fit well with a simple power-law with
$\Gamma=1.9$ and a small amount of intrinsic absorption.  The
{\it XMM-Newton} OM UV flux measurement is consistent with an extrapolation
of the X-ray power-law, suggesting that the UV to X-ray spectrum of
NGC 3998 is consistent with a power-law over 5 decades in energy.  The
SED of NGC 3998 is consistent with both RIAF models and a Comptonized
thin-disk model (i.e., as expected to be found in Seyfert galaxies),
both of which can account for the UV to X-ray spectrum of NGC 3998.
None of the accretion models account for the compact radio emission in
NGC 3998 which is probably dominated by a jet.  
The ``pure'' thin-disk model also does not account the
mid-IR emission, although a truncated thin-disk surrounding
an inner RIAF fits the mid-IR to X-ray spectrum well (with the near-IR
to optical region dominated by extra-nuclear flux).  
If the accretion is dominated by an RIAF then the transition radius is $\ga
300R_s$ and the $\dot{m}$ is $\sim 3 \times 10^{-3}$ (at $\sim 300R_s$ in the
models with p $>$ 0).
In either case the inner radius of the thin disk is probably $\ga 100R_s$ in
order to account for the lack of X-ray reflection features.  If the accretion
is dominated by a Comptonized thin disk that extends to a radius of
3$R_s$, then the accretion rate in the disk is in the range $10^{-6} -
10^{-5}$ in Eddington units and the
lack of Fe-K emission may be due to an ionized accretion
disk. Finally, a jet model may be consistent with the radio and
X-ray flux, and jet emission may also be responsible for the mid-IR
and optical/UV flux.

\acknowledgments 
This work was supported by NASA grants NAG5-11378 and NAG5-11572.  
Y. T. is supported by the Japan Society for the Promotion of Science.
E. Q. received support by NASA ATP grant NAG5-12043 and a Sloan Foundation
Fellowship. We made use of the High Energy Archive and Research Center at
NASA/GSFC and the NASA Extragalactic Database.  We thank the anonymous
referee for useful comments that improved this paper.

\end{document}